\documentstyle[12pt]{article}
\begin{document}
\setlength{\baselineskip}{24pt}

\begin{center}
{\large SYMMETRIZED PQMC STUDIES OF THE GROUND STATE OF $C_{60}$}
\footnote{Contribution no. 1169   from the Solid State
and Structural Chemistry Unit } \\
\vspace{1cm}
{Bhargavi Srinivasan$^{3}$, S. Ramasesha$^{3,5}$ and H. R.
Krishnamurthy$^{4,5}$  }
{\footnote{e-mail: bhargavi@sscu.iisc.ernet.in,~ramasesh@sscu.iisc.ernet.in,\\
{~~~~~~hrkrish@physics.iisc.ernet.in}} }\\
\vspace{0.5cm}
{\it
$^{3}$Solid State and Structural Chemistry Unit  \\
Indian Institute of Science, Bangalore 560 012, India \\
\vspace{0.5cm}
$^{4}$Department of Physics  \\
Indian Institute of Science, Bangalore 560 012, India \\   
\vspace{0.5cm} 
$^{5}$Jawaharlal Nehru Centre for Advanced Scientific Research \\
Jakkur Campus, Bangalore 560 064, India \\
}
\end{center}
\vspace{1cm}

\pagebreak
\clearpage
\begin{center}
{\bf \underline{Abstract}}\\
\end{center}
We develop a new $\it {symmetrized}$ version of the Projector Quantum Monte 
Carlo (SPQMC) method 
which preserves the symmetries of the system 
by simultaneously sampling all symmetry related Ising configurations 
at each MC step
and use it to study the effect of electron 
correlations on a single $C_{60}$
molecule  and its structural motifs, within the Hubbard model. 
Thi SPQMC method allows more accurate estimates of correlation
functions as seen from calculations on small systems.  
The method applied to some molecular fragments of $C_{60}$, viz,
pyracylene, fluoranthene and corannulene, gives a good chemical
description of these systems.  Analysis of the ground state bond 
orders allows us to visualize pyracylene  as  a 
naphthalene  moiety with  weakly bridged ethylenic units, 
fluoranthene as weakly bridged naphthalene and benzene units,
and corannulene as radialene with ethylenic bridges. We study
the ground state properties of a single  fullerene molecule, with and
without bond alternation.  The bond orders in the ground state 
for the two types of nearest neighbour bonds are unequal,
even for uniform $C_{60}$.
The spin and charge correlations give a consistent
picture of the interacting ground state  in all these systems.
We construct projections of  the bond orders and 
spin-spin correlation functions  on the space of irreducible representations
 of the icosahedral point group. These projections,  analogous
 to the structure factors of translationally invariant systems,  give
the  amplitudes for distortions which transform as the irreducible
representations of the point group. The amplitude for the $H_{g}$ distortion
is the largest while the spin structure  has large weights in the
$T_{2g}$ and $G_{u}$ representations.   
\pagebreak
\section{Introduction}

The fascinating and esoteric phenomena exhibited by systems
containing the  fullerene, $C_{60}$\cite{kroto,krats}, have attracted the
attention of experimentalists and theorists in both physics and
chemistry.  Pure $C_{60}$ in the soild state exhibits interesting
plastic crystal transformations\cite{tycko}.  Alkali doped compounds of $C_{60}$
are superconducting, with $Rb_{1.5}Cs_{1.5}C_{60}$  having the
highest known  
superconducting transition temperature for organic systems ($T_{c}$
~= 32K)\cite{fleming}. The donor--acceptor compound,  $C_{60}-TDAE$
($TDAE$ = tetrakis dimethyl amino ethylene) is one of the few known purely
 organic ferromagnets, with a  Curie temperature of 16.5K, 
the highest known for an organic system\cite{alle,steph}. 

Despite intense efforts at understanding the diverse electronic
phenomena in $C_{60}$ systems,  a reliable correlated electronic
structure for the ground state of even an isolated $C_{60}$ molecule still
remains elusive.  A perturbation theoretic  (PT) calculation in the
 framework of the Hubbard and modified
Hubbard models\cite{sudip,sondhi} focuses on the energies of the 
low--lying states of doped $C_{60}$. The  main concern of this study 
is the sign of the
low--spin high--spin energy gap as a function 
of the model parameters.  This problem has also been studied by the Variational 
Monte Carlo (VMC) method\cite{kriv} using a Gutzwiller type variational 
wavefunction\cite{gutz}.
The large $U$ limit  of the Hubbard model on a single fullerene molecule leads
to a frustrated, antiferromagnetically interacting 
 $s=1/2$  Heisenberg spin model.  The classical ground state
(large $s$ limit) has been determined\cite{coffey}. Using
a spin--wave analysis around the classical ground state,  the
possibilty   of a  magnetic instability in the ground state at a critical value $U_{c}$
of  the Hubbard parameter has been conjectured\cite{joli}.  

 In recent years, there have been some reports of
calculations of the ground state of $C_{60}$ using conventional 
quantum chemical techniques.
One of these employs  an $STO-3G^{*}$ basis in the Hartree--Fock
procedure and follows this up with a limited configuration
interaction (CI) procedure\cite{schulman}.  Another calculation 
employs the Modified
Neglect of Differential Overlap (MNDO) procedure\cite{bako} 
while yet another the Linearized Muffin Tin Orbital  (LMTO) method\cite{
sath}.
Even at the level of a single fullerene molecule, conventional quantum
chemical techniques for  obtaining the correlated electronic
structure are inadequate. 
  Besides being basis set dependent,
 these approximations are known to be uncontrolled due to arbitrary truncation of the
CI procedure. Thus, such
calculations fail to isolate the role of the active $p_{\pi}$-type orbital
in determining the properties of fullerene.  Given these
difficulties encountered in  quantum chemical approaches,
what is essential to understanding the properties of $C_{60}$ is a
thorough study of well defined model hamiltonians. Such studies allow
us to follow the evolution of the electronic states as a function of
model parameters, thereby providing valuable insights. 

The solution of model hamiltonians  can also be undertaken within
the framework of quantum chemical techniques, but even here, the
astronomically large dimensionality of the full many body space  corresponding
to the half filled (or nearly  half filled) $p_{\pi}$-like orbitals
cannot be dealt with reasonably. However, exact calculations are
possible on chemically feasible fragments of $C_{60}$
and these studies would provide the necessary  calibration for any 
novel approximate technique\cite{bssrssc,acyam}.

Amongst recent many body techniques for solving interacting model
hamiltonians,  Quantum Monte  Carlo (QMC) methods and the density matrix 
renormalization group (DMRG) method\cite{white}  have come to the fore. The DMRG
method is accurate mainly for one-dimensional lattices. There are
many variations of the QMC method.  While the VMC method\cite{yok} provides
an estimate of expectation values of the Hubbard Hamiltonian in the
Gutzwiller trial function, the temperature QMC methods\cite{temp} extrapolate
ground state properties from finite temperature properties.
A  Quantum Monte Carlo procedure that directly estimates the 
ground state properties of Hubbard type models is the Projector Quantum
Monte Carlo (PQMC) method\cite{sor}. In this method the ground state is 
projected from  a non--interacting trial  wavefunction such as
a Slater Determinant in the Huckel Molecular Orbital (MO) basis. 
Also, the PQMC method does not suffer from  either the limitations 
of the structure imposed by a trial wavefunction, as in the VMC method,
or the necessity of extrapolating the ground state from  finite temperature 
results, as in the temperature Monte Carlo methods
(which suffer from numerical difficulties at  very low temperatures)\cite{lowt}. 
 
  In this article, we report  results of the first PQMC studies of 
  the ground state of an isolated fullerene molecule.  We have greatly enhanced the
  accuracy of the Monte Carlo sampling by exploiting the symmetries
of the system. From comparison of exact results with symmetrized 
PQMC results for small clusters, we find that incorporating symmetries
leads to a marked improvement in the correlation functions, thus
providing us a more reliable picture of the correlated ground state
of $C_{60}$.  

This article is organized as follows. In the next section we 
describe  the Projector Quantum Monte Carlo method for the Hubbard 
model. In section 3, 
we introduce the symmetrized PQMC (SPQMC) method and compare results for small
systems with results obtained from exact calculations and
the VMC method.  Section 4 pertains to our results on the ground state
properties of $C_{60}$ and its molecular fragments and discussion of the data. Section 5 is a
summary of the work presented in this paper.

\section{The PQMC Method}

The Monte Carlo algorithm to project out the ground state from a
trial wavefunction was first introduced by Sorella {\it et al}\cite{sor}.
 While their
method used a fast Fourier transform (FFT) technique for propagating the single--particle
states and Langevin dynamics to sample the distribution, Imada and
Hatsugai\cite{imada} introduced an algorithm that runs parallel to the
temperature QMC, which we employ in our calculations.

The single band  Hubbard Hamiltonian, $\hat{H}$ may be written as a sum of
a non-interacting part, $\hat{H}_{0}$ and an interaction part, $\hat{H}_{1}$,
given by\cite{hub}, 
\begin{equation}
\hat{H} = \hat{H}_{0} + \hat{H}_{1} 
 \end{equation}
 \begin{equation}
  \hat{H}_{0} =  \sum_{<ij>}\sum_{\sigma}t_{ij}
(c_{i\sigma}^{\dagger}c_{j\sigma}+h.c.)
\end{equation}
\begin{equation}
 \hat{H}_{1} = U\sum_{i}\hat{n}_{i \uparrow}\hat{n}_{i \downarrow}\
\end{equation}
\noindent          
where $c^{\dagger}_{i\sigma}$ ($c_{i\sigma}$) is the
creation(annihilation) operator for an
electron with spin ${\sigma}$ in the Wannier orbital at the $i^{th}$ 
site, $\hat{n}$$_{i\sigma}$ are the corresponding  number operators
 and the summation is over bonded atom pairs.

The ground state, $|\psi_{0}>$  of a Hamiltonian, $\hat{H}$ can be projected from a
trial wavefunction $|\phi>$  by the {\it ansatz}
\begin{equation}
|\psi_{0}> = \lim\limits_{\beta \rightarrow \infty} { {e^{- \beta{\hat H}}|\phi>}
 \over {\sqrt{<\phi| e^{- 2 \beta{\hat H}} |\phi>}}}
\end{equation}
\noindent
provided $|\phi>$ has a nonzero projection on to $|\psi_{0}>$. 
The trial wavefunction $|\phi>$ is usually formed from the MOs generated by
diagonalizing the non-interacting problem. The trial state with
$M$-fermions, in second--quantization is represented as 
\begin{equation}
|\phi_{\sigma}> = \prod\limits_{m=1}^{M} \Bigl( \sum\limits_{i=1}^{N}
({\bf{\Phi}}_{\sigma})_{im} \hat{c}_{i\sigma}^{\dagger} \Bigr) |0>
\end{equation}
\noindent 
where ${\bf{\Phi}}_{\sigma}$ is an $N \times M$ matrix of  the MO
coefficients  whose row index, $i$, corresponds to the site number and column
index,  $m$, corresponds to  the $m^{th}$ MO  occupied by an electron of
spin $\sigma$, in the trial wavefunction.

To carry out the 
projection for an interacting Hamiltonian, it is necessary to Trotter 
decompose the density operator $exp(-\beta \hat{H})$ as 
\begin{equation}
e^{- \beta {\hat H}} =  \Bigl( e^{- \Delta \tau {\hat H}} {\Bigr)}^{L}
\end{equation}
\noindent
\begin{equation}
 e^{- \Delta \tau{\hat H}} = e^{- \Delta \tau /2 {\hat H_{0}} }
e^{- \Delta \tau {\hat H_{1}} }   e^{- \Delta \tau /2 {\hat H_{0}} }
\end{equation}
\noindent
where $\beta = L \times \Delta\tau$. The error introduced in this decomposition due to
noncommutativity of $\hat{H}_{0}$ and $\hat{H}_{1}$ is of order $(\Delta\tau)^{3}$.
 The matrix representation,  ${\bf{b}}_{0}$, of
$exp(-\Delta\tau/2 \hat{H_{0}})$  can be easily obtained in the basis of
Wannier functions as
\begin{equation}
 {\bf{b}}_{0} = exp [ - {\bf{K}}]
\end{equation}
\noindent
\begin{equation}
   K_{ij} = \cases{  -{{\Delta\tau\over2} t_{ij}}, & for $i,j$ bonded sites; \cr
                           0, & otherwise.}
\end{equation}
\noindent

However, the interaction part cannot be represented as a matrix in the
one-particle basis. This makes it necessary to obtain a one-particle 
like description of $exp(\Delta \tau \hat{H}_{1})$, which is indeed possible via a discrete
Hubbard-Stratanovich (H-S) transformation at every time slice. At the $l^{th}$ time slice,
the H-S transformation gives
\begin{equation}
e^{- \Delta\tau \hat{H}_{1}(l) } = 
\sum\limits_{s_{1l}=\pm 1} \sum\limits_{s_{2l}=\pm 1} \ldots
\sum\limits_{s_{Nl}=\pm 1}
 e^{- \Delta\tau \hat{H}_{1 \uparrow}(l,{\bf{s}}_{l})}
 e^{-\Delta\tau \hat{H}_{1 \downarrow}(l,{\bf{s}}_{l})}
\end{equation}
\begin{equation}
e^{- \Delta\tau \hat{H}_{1 \uparrow}(l,{\bf{s}}_{l})} =
 \prod\limits_{i=1}^{N} \Bigl( {1\over 2}  exp (\lambda s_{il}\hat{n}_{i\uparrow}
  - \Delta\tau {U \over 2}\hat{n}_{i \uparrow}) \Bigr) 
\end{equation}
\begin{equation}
e^{- \Delta\tau \hat{H}_{1 \downarrow}(l,{\bf{s}}_{l})} =
 \prod\limits_{i=1}^{N}\Bigl( {1\over 2}  exp (-\lambda s_{il}\hat{n}_{i\downarrow}
 - \Delta\tau {U \over 2}\hat{n}_{i \downarrow}) \Bigr)
\end{equation}
\noindent
where 
\begin{equation}
\lambda = 2 arctanh \sqrt{tanh(\Delta\tau U/4)},
\end{equation}
\noindent
$s_{il}$ is the Hubbard-Stratanovich field at the $i^{th}$
site in the $l^{th}$ time slice and ${\bf{s}}_{l}$ is an $N$-vector whose
$i^{th}$ component corresponds to the Ising  variable $s_{il}$. Now 
it becomes possible to 
obtain a matrix representation, ${\bf{b}}_{1 \sigma}(l,{\bf{s}}_{l})$, 
of $exp(-\Delta\tau H_{1 \sigma}(l,{\bf{s}}_{l}))$
 in the basis of Wannier functions,
\begin{equation}
  ({\bf{b}}_{1\uparrow}(l,{\bf{s}}_{l}))_{ij} = \delta_{ij} {1\over2} 
  exp [\lambda s_{il} - {\Delta\tau \over 2} ]
\end{equation}
\begin{equation}
  ( {\bf{b}}_{1\downarrow}(l,{\bf{s}}_{l}))_{ij} = \delta_{ij} {1\over2} 
  exp [-\lambda s_{il} - {\Delta\tau \over 2} ]  .
\end{equation}
\noindent
This gives us the matrix representation, ${\bf{B}}_{\sigma}(l,{\bf{s}}_{l})$,
for  $exp(- \Delta \tau{\hat{H}_{\sigma}})$ at the $l^{th}$ time slice,  
\begin{equation}
{\bf{B}}_{\sigma}(l,{\bf{s}}_{l}) = {\bf{b}}_{0}  {\bf{b}}_{1 \sigma}
(l,{\bf{s}}_{l})  {\bf{b}}_{0}
\end{equation}
\noindent

The Trotter decomposition followed by the H-S transformation thus  results
in the replacement of the 
interacting Hamiltonian by a sum of $2^{N \times L}$ noninteracting
 Hamiltonians  corresponding to each configuration of the  $(d+1)$-dimensional
$N \times L$ Ising lattice.
The properties of the interacting Hamiltonian are simple averages taken over 
the corresponding properties of all the  $2^{N \times L}$ 
noninteracting Hamiltonians.  The matrix representation of the projection 
operator for the interacting Hamiltonian  can now be written as a sum  
of individual projections, ${\bf{P}}(\{s\})$, each corresponding to the 
non-interacting Hamiltonian for the Ising  configuration $\{s\}$. The matrix
${\bf{P}}(\{s\})$ is given by,
\begin{equation}
{\bf{P}}(\{s\}) = {\bf{P}}_{\uparrow}(\{s\}) 
{\bf{P}}_{\downarrow}(\{s\})
\end{equation}
\begin{equation}
{\bf{P}}_{\sigma}(\{s\}) = {\bf{B}}_{\sigma}(1,{\bf{s}}_{1}) 
{\bf{B}}_{\sigma}(2,{\bf{s}}_{2})\ldots
{\bf{B}}_{\sigma}(L,{\bf{s}}_{L}) 
\end{equation}
\noindent
This allows  the representation of the expectation value, $w$,  of the density 
operator $exp(-\beta \hat{H})$, in the trial state $|\phi>$, namely,
\begin{equation}
w  =  \langle \phi |  e^{- \beta \hat{H}} | \phi  \rangle
\end{equation}
\noindent
which takes the form 
\begin{equation}
w  =  \langle \phi |\sum\limits_{\{s\}}  \bf{P}(\{s\}) | \phi \rangle
\end{equation}
\noindent
within Trotter error. It is convenient to define  weights for each
Ising configuration $\{s\}$ as 
\begin{equation}
 W ( \{s\} ) =  W_{\uparrow} ( \{s\} )
 W_{\downarrow} ( \{s\} )
 \end{equation}
 \begin{equation}
W_{\sigma}(\{s\}) = det {\Bigl(} {\bf{\Phi}}_{\sigma}^{T} {\bf{B}}_{\sigma}
(1,{\bf{s}}_{1})  {\bf{B}}_{\sigma}(2,{\bf{s}}_{2})\ldots
{\bf{B}}_{\sigma}(L,{\bf{s}}_{L} )  {\bf{\Phi}}_{\sigma} {\Bigr)}
\end{equation}
\noindent
$w$ can now be conveniently expressed as a sum over weights of all
Ising configurations, 
\begin{equation}
 w  = \sum\limits_{\{s\}} W ( \{s\}).
\end{equation}

For large $N$ and $L$, it is computationally prohibitive  to obtain 
properties as exact averages over all Ising configurations.
 A simple average based on crude sampling would lead to large errors
in the estimates. However, by resorting to importance sampling of the Ising
configurations, it is possible to obtain estimates with acceptable errors.
A straightforward importance sampling of the Ising configurations assumes 
that an Ising configuration $\{s\}$ appears with  probability
\begin{equation}
   \pi(\{s\})  =  {{ W( \{s\} ) } \over { w}}
\end{equation}
 However, computing $w$ is equivalent to obtaining the ground
state projection for the given $\beta$. This vicious circle is broken by employing
an ergodic Markov chain.

 The state space of the Markov chain is taken to be the 
space of Ising configurations and the initial Ising configuration is chosen 
assuming a uniform probability distribution in the configuration space. The 
one-step transition probabilities between any two Ising configurations $\{s\}$ and $\{s'\}$
is so chosen that the ratio of the limiting probabilities of the configurations tends 
to ${\pi(\{s^{\prime}\}) \over \pi(\{s\}}$. The one-step transition probability 
\begin{equation}
  p_{\{s\}\rightarrow\{s\prime\}} = { {r} \over {1+r}}
\end{equation}
\begin{equation}
  r = {  { W(\{s^{\prime}\}) } \over { W(\{s\}) }  }
\end{equation}
\noindent
gives the correct limiting ratios for the probabilities of Ising configurations.
The Monte Carlo simulation now follows the usual procedure. The new Ising
configurations can be generated either by the cluster spin-flip or the single
spin-flip mechanisms. 
For convenience we employ sequential spin flips exhausting all sites in a
given time slice before moving over to the next time slice.

 In the Monte Carlo
procedure, we estimate the equal time Green function
$G_{\sigma}(l,\{s\})$, which we regard as  a matrix in the Wannier basis,
for spin $\sigma$, at the $l^{th}$ time slice, for every configuration $\{s\}$
which is sampled. The $(j,k)^{th}$ matrix element of
$G_{\sigma}(l,\{s\})$  is given by
\begin{equation}
(G_{\sigma}(l,\{s\}))_{jk} =  {  \langle L_{\sigma}(l,\{s\})  \hat{c}_{k \sigma}^{\dagger}
\hat{c}_{j \sigma} | R_{\sigma}(l,\{s\}) \rangle \over
\langle L_{\sigma}(l,\{s\}) | R_{\sigma}(l,\{s\}) \rangle }
\end{equation}
\noindent
where the  the projected states  $\langle L_{\sigma} (l,\{s\}) |$  and 
$ |R_{\sigma} (l,\{s\}) \rangle$  are obtained
analogous to $|\phi>$ in eqn. (5) using the coefficient matrices  ${\bf{L}}$ and ${\bf{R}}$,
\begin{equation}
{\bf{L}}_{\sigma}(l,\{s\}) = {\bf{\Phi}}_{\sigma}^{T}  {\bf{B}}_{\sigma}(1,{\bf{s}}_{1}) 
{\bf{B}}_{\sigma}(2,{\bf{s}}_{2}) \ldots
{\bf{B}}_{\sigma}(l-1,{\bf{s}}_{l-1})  {\bf{b}}_{0}
\end{equation}
\begin{equation}
{\bf{R}}_{\sigma}(l,\{s\}) =  {\bf{b}}_{1 \sigma}(l,{\bf{s}}_{l})  {\bf{b}}_{0} 
{\bf{B}}_{\sigma}(l+1,{\bf{s}}_{l+1})
{\bf{B}}_{\sigma}(l+2,{\bf{s}}_{l+2}) 
\ldots
{\bf{B}}_{\sigma}(L,{\bf{s}}_{L}) {\bf{\Phi}}_{\sigma}.
\end{equation}
In practice, this involves obtaining the inverse of a matrix. For the single spin
flip mechanism, efficient algorithms exist to obtain the ratio, $r$, of the
determinants and to update the  Green function
if the spin flip is accepted\cite{imada,numrec}. However, the updated 
inverse degrades fairly
rapidly and the Green function will have to be recomputed from scratch
every few hundred spin flips. Properties of the Hamiltonian can be evaluated 
from  Monte Carlo estimates of the equal time Green function  
$\langle\langle G \rangle \rangle$ 
\begin{equation}
\langle\langle G_{\sigma} \rangle \rangle_{ij} = {{1} \over {N_{s}LN}}
\sum\limits_{l,\{s\}}  (G_{\sigma}(l, \{s\})_{ij}.
\end{equation}
\noindent
where $N_{s}$ is the number of MC sweeps over the Ising lattice,
after equilibration.
We employ Wick's theorem to compute two particle properties from the
single-particle Green function, $(G_{\sigma}(l, \{s\}))_{ij}$, at every time
slice, $l$. 

\section{Symmetrized PQMC Method}

A serious drawback of Monte Carlo simulations carried out as
described in the previous section is that the  Green function
and correlation functions do not  reflect the  symmetries of the
system. This is because the noninteracting Hamiltonian corresponding
to an Ising configuration does not have the full symmetry of the
system. In the MC sampling  each Ising configuration is dealt with
independently. Thus, when a particular  Ising configuration is
visited in the course of sampling, it is not guaranteed that all the other
Ising configurations related  by symmetry are also visited.
For example, while we may sample the configuration in fig. 1a, it
is not certain that the
configuration in fig. 1b, which is related by $C_{4}$ symmetry is also
sampled during the simulation. For this reason, the MC sampling breaks
the  symmetry of the Hamiltonian. The full symmetry of the 
Hamiltonian can be preserved if we ensure that for every  Ising 
configuration that is sampled, all its symmetry related partners 
are also tacitly included in the sampling.  

The symmetry of the Hamiltonian guarantees that the one--step
transition probability between Ising configurations $\{s\}$,
$\{s^{\prime}\}$ is the same as that between  $\hat{R}\{s\}$ and
$\hat{R}\{s^{\prime}\}$, where $\hat{R}$ is a symmetry element of the
Schrodinger group of the Hamiltonian,
\begin{equation}
p_{\{s\}\rightarrow\{s^{\prime}\}} =  p_{\hat{R}\{s\}\rightarrow\hat{R}
\{s^{\prime}\}}
\end{equation}
\noindent
Therefore, if an Ising configuration $\{s^{\prime}\}$ is accepted (rejected) 
from an
initial configuration $\{s\}$, the same result is expected from all symmetry
related pairs of configurations $\hat{R}\{s^{\prime}\}$ and $\hat{R}\{s\}$.
This feature can be incorporated by  constructing a symmetrized Green
function as follows:
\begin{equation}
G^{sym}_{ij}(\{s\})   =   {1\over h} \sum\limits_{ {\hat{R}}}
G_{ij}({\hat{R}}\{s\})   
\end{equation}
\noindent
and $\hat{R}$ runs over all the $h$ symmetry elements of the group.

It appears from the equation that we need to update the Green functions
$G(\hat{R}\{s\})$ for every symmetry operation $\hat{R}$, which could 
be enormously compute intensive. However, the Green functions
 $G(\hat{R}\{s\})$ and $G(\{s\})$ are related as
\begin{equation}
G_{ij}(\{s^{\prime}\})   =   G_{i^{\prime}j^{\prime}}(\{s\})   
\end{equation}
\begin{equation}
\hat{R} :   \{s\}  \rightarrow \{s^{\prime}\};
\end{equation}
\begin{equation}
   \hat{R}^{-1} :  i \rightarrow  i^{\prime} \hspace{0.2cm}  ; 
\hspace{0.2cm}   \hat{R}^{-1} :  j \rightarrow  j^{\prime}
\end{equation}
\noindent
Thus, from the Green function of a  single Ising configuration, we can generate
the Green functions of all Ising configurations related by the symmetry
group of the system.

In  highly symmetric systems such as $C_{60}$, the number of 
symmetry elements, $h$, of the Schrodinger group  can be very large.  
For instance,  the number of symmetry elements in the icosahedral group
is 120 and the number of pairs of sites is  $O( N^{2})$,  which,
for $C_{60}$ is 3600.  In a symmetrized sampling procedure, for $C_{60}$,
it appears that we should average  all the 3600 elements of the single particle
Green function over the 120 symmetry related Ising configurations. 
This procedure turns out to be prohibitive, computationally.  However,
we recognise that there are only 24 unique pairs of carbon atoms on the 
fullerene molecule.  It is sufficient to average these 24 pairs of  matrix 
elements of the Green function over all the  symmetry related Ising
configurations, to obtain the  complete symmetrized Green function. 
This reduces the cpu requirement, in the case of
$C_{60}$ by a  factor of 150,  thus rendering the SPQMC 
procedure computationally feasible for large systems.  

The PQMC method in its current  implementation is not variational. A variational 
implementation of the PQMC procedure can, in principle, be acheived by 
projecting the states $\langle L_{\sigma}(l,\{s\})|$ and 
$|R_{\sigma}(l,\{s\}) \rangle$ from identical Ising
configurations. The Green function is sampled only at the middle
time slice. This ensures that the estimated quantity is an expectation
value of a projected trial state. However, this version of PQMC would have
larger errors for the same $\beta$ due to halving of the Ising lattice.

We have tested the SPQMC procedure for the simple case of a Hubbard ring 
of six sites since this is easily amenable to exact studies. In table 1,
we compare the  SPQMC  ground state energies for different values of the Hubbard 
parameter, $U/t$, with the  results of  exact and VMC calculations. It is
not surprising that  in terms of energy, the VMC results are better than
the PQMC results for very small $U (\approx 0.5t)$, since the Gutzwiller trial
function is the true ground state for $U/t=0.0$.  However,  for  larger 
values of $U/t$, the percentage errors in PQMC estimates of the ground state
energy are smaller than in the VMC estimates.  The most striking 
difference, however, is seen in  two-particle
correlation functions  like the spin-spin  and charge-charge  correlations. 
The  results presented in tables 2a and 2b show clearly that the
SPQMC results are accurate even for $U=4.0t$.  Thus, the SPQMC method allows
a better understanding of the  correlated ground state of  $C_{60}$.

Symmetrized sampling can also be used to estimate properties of 
excited states. This is possible since the projection operator
projects the lowest energy state in a chosen symmetry subspace
of the Hamiltonian. We could start with an initial state that 
transforms as the basis of an irreducible representation of the
desired  space, usually  an appropriate 
linear combination of Slater Determinants obtained from  group theoretic 
projection operators. In the symmetrized sampling procedure, the  
symmetry of the Hamiltonian is explicitly preserved. This prevents
any admixture of states from different symmetry spaces thereby 
leading to the lowest energy state of the subspace. When the  starting 
state is a linear combination of Slater Determinants,  estimates of 
properties require modifications to the PQMC procedure as will be 
discussed elsewhere\cite{mltdet}. 

For simple cases with non-degenerate LUMOs, such as Hubbard chains 
the properties of the ground state of the doped system can be obtained
from a single determinantal PQMC with symmetrized sampling,
as verified for small systems\cite{mltdet}. The doped states of molecules with degenerate
LUMOs are rather difficult to handle. To estimate the ground state 
properties of such systems, with open shell electronic 
configurations, it is necessary to carry out a multi-determinantal,
symmetrized PQMC simulation. This is true for even as simple a
system as doped benzene. There is considerable interest in
the  electronic structure of doped $C_{60}$ species. Unfortunately,
the threefold degeneracy of the LUMOs of $C_{60}$ makes a single determinantal
calculation feasible only for $C_{60}^{6-}$ and the high spin
state of $C_{60}^{3-}$. We are currently implementing a
multideterminantal, symmetrized PQMC method for studying doped
$C_{60}$ species.

\section{Results on the Ground State of $C_{60}$ and its Fragments}

In this section we present results on the ground state of 
$C_{60}$ and its fragments, obtained via SPQMC calculations. We compare
these with VMC results and also with exact results, wherever
possible\cite{acyam,bssrhrk1}. In Section (4.1) we discuss the nature of the correlated
ground state in the molecules pyracylene, fluoranthene and corannulene 
(fig. 2). These molecules have been chosen because they represent 
the characteristic
chemical motifs of $C_{60}$ and  contain one or more
five--memebered rings  essential for displaying the minimal
topological features of $C_{60}$. The trends observed in these systems
are expected to be indicative of the behaviour of  the full molecule.
In Section 4.2 we present SPQMC results on both uniform and bond-alternated
$C_{60}$ for various values of $U/t$. These results include energies, bond
orders, charge and spin correlations. We compare these results with the VMC
results.

\subsection{Fragments of $C_{60}$}
In table 3 we present the ground state energies of the fragments, pyracylene,
fluoranthene and corannulene for different values of $U/t$.
The pyracylene molecule is the smallest characteristic fragment of $C_{60}$ 
that we have studied. It is also amenable to exact calculations. 
 The ground state energy obtained from the SPQMC calculations 
compares  favourably with that from the exact calculation for $U/t=1.0$.
We also compare our SPQMC results with the VMC results for all 
values of $U/t$, for all fragments. The agreement is very good with VMC for
small values of $U/t$ and tends to deteriorate for larger values of the Hubbard
parameter. The SPQMC results are more reliable than the corresponding VMC
results, for similar sample sizes. For, the SPQMC  estimates are exact
within Trotter error (which in our case is $O(10^{-3})$). In contrast, the VMC 
estimates are also limited by the choice of the trial wavefunction and
it has been shown that the Gutzwiller function,
with a single variational parameter, is a poor trial function at large $U/t$
\cite{kaplan}.

The bond orders  in pyracylene, fluoranthene and 
corannulene decrease with increasing $U/t$,
which is consistent with increased covalency of the ground state. A detailed
analysis of the bond order data allows one to visualize these molecules
in terms of interconnected chemical motifs. We can describe pyracylene as
a naphthalene moiety weakly coupled to two ethylene moities. Fluoranthene can
be visualized as a naphthalene motif which is weakly linked to a benzene 
unit. And corannulene can be viewed as ethylene bridges connecting
the spokes of the radialene molecule.

The spin correlations of the nearest neighbours in these molecules
 lend support to the picture emerging
from  the bond order data.  The  strong  bonds  also have  large
antiferromagnetic ({\it afm}) correlations. Thus, the strongest  bond
(1-2 bond)  in pyracylene also has  the largest {\it afm} spin-spin 
correlation.  In fluoranthene, the bridging  bond (4-5 bond) has a spin-spin
correlation which is less than  a third of that of the strongest bond 
(7-8 bond).
The bonds along the perimeter  of corannulene fall into 
two distinct classes, a trend expected from the bond order variations.
What is more interesting are the bonds of the radialene motif (fig. 3),
all of which have reasonably large  {\it afm} correlations. 
However the spokes have  larger correlations than the bonds of 
the pentagon. Thus we see that the  pentagon in corannulene 
is already in a chemical environment similar to that of $C_{60}$.
Frustration  due to the five membered ring is  thus
manifest in a qualitatively different way, in this molecule. 

 The charge correlations of the bonds in pyracylene, fluoranthene and corannulene 
increase with increasing $U/t$, while the
on-site (diagonal) charge correlations decrease with increasing $U/t$, 
a feature expected of the Hubbard model. In the large $U/t$ limit,
the intersite charge correlations approach unity. The magnitude
of the charge correlations can be related to the total oscillator
strength of the molecule for dipole transitions\cite{osc}. This evidently
decreases with increase in the strength of electron correlations.
The diagonal charge correlations in the noninteracting limit have 
a value of 1.5 while in the strongly interacting case, this valuse is 1.0.
However, depending upon the topology of the bond connectivities, the 
interactions  lead to values of the diagonal charge correlations 
in the range 0--2. The  picture as obtained from charge 
correlations computed using the SPQMC  method corresponds to a more uniform 
distribution of  charges than that obtained from the VMC calculations.

\subsection{SPQMC Ground State Properties of $C_{60}$}

We have computed the ground state properties of an isolated fullerene molecule using 
the SPQMC method. We have studied  uniform fullerene (equal
hexagon-hexagon (h-h) and  hexagon-pentagon (h-p) transfer integrals) 
as well as  bond alternated fullerene (h-h transfer integral
larger in magnitude than the h-p transfer integral).  In the 
latter, the  h-h and h-p bonds were taken to have 
transfer matrix elements $t_{h-h}$ and $t_{h-p}$  equal to
1.2 and 0.9 respectively, so that the sum of the transfer integrals
remains unchanged after introducing alternation.  This ratio, which is slightly larger than the
experimental value, was chosen to magnify the effects of bond alternation and
is not expected to change the properties qualitatively.  We have carried out
the simulations with $\Delta \tau = 0.1$,  usually  for a 
value of the projection parameter, $\beta \approx 2.0$. This 
corresponds to a Trotter error of $\approx {10^{-3}}$. We have employed
$\approx {10^4}$ MC sweeps, allowing 2000 sweeps 
for equilibration. We have checked for convergence by increasing the value of
the projection parameter $\beta$, number of MC sweeps and decreasing 
$\Delta\tau$ and found the above parameters to be satisfactory.

In table 4 we present the ground state energy, energy/site 
and energy/bond of
$C_{60}$ with and without bond alternation. Compared to its 
fragments,  $C_{60}$ has the highest energy/bond
at any give value of $U/t$, indicating that it is the most frustrated
of the systems studied. The ground state energy increases with 
increasing interaction strength but decreases upon introducing bond
alternation. We  compare  the ground state energies obtained from 
the SPQMC calculations with those from VMC studies in fig. 4
\cite{kriv,bssrhrk1}. 
Since the SPQMC calculations are not variationally bounded caution
should be exercised in interpreting the PQMC energies {\it vis \`a vis} the VMC
energies. 
However, test calculations on small systems indicate that at
intermediate $U/t$, the absolute SPQMC errors in energies are comparable
with or smaller than VMC errors. We shall presently see that the description
of the ground state within the SPQMC calculation is better than that
afforded by the VMC method.

$C_{60}$ perfers a bond alternating structure in the non-interacting
limit since even here the h-h bonds have larger bond orders than the
h-p bonds\cite{wifd,hed}. Using Coulson's relation this translates 
into shorter h-h distances compared to h-p distances\cite{coul,mall}. 
In table 5 we present the bond orders of uniform and bond alternating
$C_{60}$ for different values of the Hubbard parameter.  We also compare
these values with the VMC bond orders in fig 5. The
VMC bond orders  show a stronger dependence on $U/t$. In table 6 we
present the difference in the bond orders of the h-h and the h-p
bonds for uniform and bond alternant $C_{60}$, within VMC and SPQMC schemes.
 The VMC calculations seem to indicate that interactions
supress this difference.  However, the SPQMC bond orders  do not 
fall so sharply with increasing $U/t$.  In one-
dimensional Peierls-Hubbard systems, it is known that the difference 
in bond orders of the '{\it double}' and '{\it single}' bonds goes through
a maximum as a function of $U/t$\cite{srpeierls}.  It is reasonble to expect 
a similar behaviour in the case of $C_{60}$.  However, the maximum in
this case should shift to larger values of $U/t$ due to increased
band width of the bands in higher dimensions.  The SPQMC bond order
difference registers a downward  trend for $U/t$ between 5.0 and 6.0.

We examine the spin-spin correlations on  fullerene for different
interaction strengths (table 5, fig. 6) and for different
inter-site separations (fig. 7) for connectivities shown in
the pentagonal projection  of $C_{60}$ shown in fig. 3.
  The h-h spin-spin correlation is 
strongly antiferromagnetic while the h-p correlation, although
{\it afm}, is weaker. This trend is already manifest even in corannulene, 
an open system.   We observe in fig. 6 the evolution  of the nearest 
neighbour and longer range correlations  with increasing $U/t$.  As with the 
bond orders, increasing interaction strength enhances the difference
between the h-h and h-p spin-spin correlations. A Hartree-Fock (HF) calculation
on $C_{60}$ predicts an instabilty in the spin structure of the
ground state for $U/t \approx 2.6$\cite{joli}.  The evolution of the SPQMC nearest
neighbor and longer range correlations with increasing correlation strength
is smooth (fig. 6), showing no indication of  any instability in the vicinity 
of $U/t \approx 3.0$. We also study the charge correlations 
(table 5 and fig. 8)  and find no evidence of any instability
in the ground state. Thus, the properties of the ground state 
tend to evolve smoothly from the non-interacting to the
strongly correlated limit. 

The SPQMC procedure allows a more detailed analysis of the bond orders
and spin-spin correlations, since all the symmetry related estimates
are strictly equal. We can construct the {\it molecular structure factors} (MSF)
of $C_{60}$,
similar to the structure factors in crystals, by projecting the quantities
such as bond orders and spin correlations on to different irreducible 
representations of the icosahedral point group. The MSF corresponding 
to the bond orders for the $\Gamma^{th}$ irreducible representation,
$\rho_{\Gamma}$, is given by 
\begin{equation}
\rho_{\Gamma} = { 1 \over h} \sum\limits_{\hat{R}}
\chi_{\Gamma}(\hat{R}) \Bigl( \sum\limits_{\sigma}
c^{\dag}_{i \sigma}  c_{\hat{R}i \sigma} + h. c.
\Bigr).
\end{equation}
\noindent
Similarly the MSF corresponding to the spin correlations $S_{\Gamma}$ is
given by
\begin{equation}
S_{\Gamma} = { 1 \over h} \sum\limits_{\hat{R}}
\chi_{\Gamma}(\hat{R}) \langle
s^{z}_{i}  s^{z}_{\hat{R}i}
\rangle
\end{equation}
where  $\chi_{\Gamma}(\hat{R})$ is the character of the symmetry operator
$\hat{R}$ in the $\Gamma^{th}$  irreducible representation of the
group and  $s^{z}_{i}  s^{z}_{\hat{R}i}$  is the spin-spin correlation 
of the $z$-component of the spins at sites $i$ and $\hat{R}i$.

In fig. 8a and 8b we show the dependence of $\rho_{\Gamma}$ on
the irreducible representation $\Gamma$ for  uniform and 
bond alternated fullerenes.  Since 
bond orders are empirically related to the bond distances, these quantities
give the amplitude for the nuclear displacement corresponding to the 
chosen representation $\Gamma$. We find that the amplitudes are very small
for the $A_{1g}$ and $A_{1u}$ representations. The former corresponds to
a breathing in and out of the fullerene molecule while the latter corresponds to
diametrically opposed motions of the nuclei related by the inversion
symmetry. The maximum amplitude is for the $H_{g}$ mode while the 
amplitudes for the other modes are almost equal in magnitude. This clearly
shows that the strongest electron-phonon coupling would be to the $H_{g}$
mode. Varma {\it et al}\cite{varma} have indeed employed electron-phonon coupling to this
mode as a mechanism for superconductivity in fullerides. The effect of
increasing the strength of electron correlations is to reduce the amplitude
of the $H_{g}$ mode. Thus, the electron-phonon coupling constant decreases 
with increase in the Hubbard parameter. Bond alternated $C_{60}$ also
shows features very similar to uniform $C_{60}$.

In fig. 9a and 9b, the dependence of the spin structure factors 
on the irreducible
representation are shown for uniform and bond alternated fullerenes. In both
the cases, we find large amplitudes for the $T_{2g}$ and $G_u$ projections.
Thus, the spin structure in both the cases are similar and correspond to 
a mix of these two representations. A geometrical picture of the spin
structure with a given $\Gamma$ can be obtained by placing individual
spins, at each site, directed along the site displacement vector for 
vibrations in that mode\cite{weeks}. It is interesting to see that the spin MSF
has an altogether different structure than the bond MSF. Thus, for instance,
the nuclear geometry resulting from application of pressure will  be
qualitatively different from the spin orientations resulting from the 
application of a magnetic field. The effect of increase in the strength 
of electron correlations on the spin structure is complex. While the amplitudes of the dominant
representations increases with increase in $U/t$, that of a few others 
decreases and this could be due to an underlying sum rule for the chosen
total spin, which in the ground state is a singlet.

\section{Summary}

We have studied the  $C_{60}$ molecule  and its molecular fragments
using a new, {\it symmetrized} version of the PQMC method developed by us. In 
this method all the symmetry related Ising configurations are 
simultaneously sampled which leads to property estimates that
reflect the full symmetry of the system. This method also
has the potential to project on to the lowest state in subspaces
of different symmetries. We have tested the SPQMC 
method on small systems. We find  a clear improvement in the 
correlation functions when compared to the nonsymmetrized estimates.

We have applied the SPQMC method to molecular fragments of $C_{60}$.
We find that corannulene, which has a pentagonal unit completely
surrounded by  ethylenic units is the fragment that resembles
$C_{60}$ most closely.  The interior sites of corannulenene 
resemble radialene which is the basic structural motif
in $C_{60}$.  We have obtained the SPQMC
ground states of $C_{60}$ with and without bond alternation. 
These have been compared with Hartree-Fock and VMC results. 
SPQMC bond orders  indicate that interactions tend to 
enhance bond alternation as in the case of one-dimensional
Peierls-Hubbard systems. We find no evidence  of an instability 
in the spin structure of $C_{60}$ with increasing correlation strength. 
We have analyzed   the bond orders and 
spin-spin correlation data in terms of their projections on
 the  space of irreducible representations of the icosahedral point group. 
These {\it molecular structure factors}  are analogous
 to the structure factors of translationally invariant systems and  give
the  amplitudes for distortions which transform as the irreducible
representations of the point group. The amplitude for the $H_{g}$ distortion
is largest and its magnitude decreases with increasing $U/t$.
Therefore, it  is expected that the $H_{g}$ vibrational mode 
 couples strongly with the electrons in the system for small 
values of $U/t$. The spin structure  
has a bimodal distribution with  large weights in the
$T_{2g}$ and $G_{u}$ representations, whose amplitudes increase
with increasing $U/t$.    

\noindent
{\bf{Acknowledgements:}} We thank Y. Anusooya for help with
 the exact computations on small systems and Dr. Biswadeb Dutta for help
with the computational facility at JNCASR.
\pagebreak
\clearpage

\begin{table}
\begin{center}
Table 1. Ground state energy of benzene (in units of $t$) as a function of $U/t$
using exact diagonalization ($E_{exact}$),  GWF as the trial wavefunction
in a Gutzwiller variational calculation ($E_{Gutz.}$), VMC and SPQMC 
calculations.  Quantities in parantheses are 
absolute  values of the percentage errors.\\
\begin{tabular} {|c| c| c| c| c|} \hline
$U/t$ & {$E_{exact}$}  & {$E_{Gutz}$} & {$E_{VMC}$}  & {$E_{SPQMC}$}  \\ \hline
0.5 & -7.274     & -7.274 (<0.007) & -7.274 (<0.007) &  -7.278 (0.05)\\ 
1.0 & -6.601     & -6.596 (0.08) & -6.599 (0.03) &  -6.611 (0.15)\\
2.0 & -5.410     & -5.386 (0.44) & -5.388 (0.41) &  -5.444 (0.63)\\
4.0 & -3.669     & -3.537 (3.60) & -3.538 (3.57) &  -3.783 (3.11)\\ \hline
\end{tabular}
\end{center}
\end{table}

\begin{table}
\begin{center}
Table 2(a). Spin--Spin correlations ($4<s_{i}^{z}s_{j}^{z}>$) 
of benzene for $U/t = 1.0$ and 4.0, from
exact and MC calculations. Quantities in parantheses are 
absolute  values of the percentage errors.\\
\begin{tabular}{|c|c|cccc|} \\  \hline
\multicolumn{1}{|c|}{$U/t$}&\multicolumn{1}{c|}{$i,j$} & \multicolumn{4}{c|}
{$4<s_{i}^{z}s_{j}^{z}>$}  \\ \hline
 & &   $Exact$  & Gutz &VMC& SPQMC \\ \hline
1.0  &1,1 &  0.567 &  0.565 (0.35) & 0.557 (1.76)  & 0.562 (0.88)  \\ 
1.0  &1,2 & -0.267 & -0.265 (0.75) & -0.245 (8.24) & -0.264 (1.12)\\
1.0  &1,3 &  0.022 &  0.016 (27.3) & 0.011 (50.0)  & 0.020 (9.10) \\
1.0  &1,4 & -0.077 & -0.068 (11.7) & -0.062 (19.5) & -0.074 (3.90)\\ \hline
4.0  &1,1 & 0.778  &  0.751 (3.47) &  0.750 (3.60) & 0.751 (3.47)\\
4.0  &1,2 &-0.435  & -0.404 (7.13) & -0.395 (9.20) & -0.412 (5.29) \\
4.0  &1,3 & 0.140  &  0.095 (32.1) &  0.077 (45.0) & 0.120 (14.0)\\
4.0  &1,4 &-0.188  & -0.131 (30.3) & -0.113 (39.9) & -0.167 (11.2)\\ \hline
\end{tabular}
\end{center}
\end{table}

\pagebreak
\clearpage

\begin{table}
\begin{center}
Table 2(b). Charge--Charge correlations  ($<n_{i}n_{j}>$) of benzene for $U/t = 1.0$
and 4.0, from
exact and MC calculations. Quantities in parantheses are the 
absolute  values of the percentage
errors. \\
\begin{tabular}{|c|c|cccc|} \\ \hline 
\multicolumn{1}{|c|}{$U/t$}&\multicolumn{1}{c|}{$i,j$} & 
 \multicolumn{4}{c|}{$<n_{i}n_{j}>$} \\ \hline 
 & &   $Exact$  & Gutz &VMC & SPQMC \\ \hline
1.0  &1,1 &  1.432  & 1.435 (0.21) & 1.443 (0.77) & 1.436 (0.28) \\ 
1.0  &1,2 &  0.816  & 0.817 (0.12) & 0.812 (0.49) & 0.813 (0.37) \\
1.0  &1,3 &  0.989  & 0.989 (<0.05) & 0.976 (1.31) & 0.989 (<0.05) \\
1.0  &1,4 &  0.960  & 0.953 (0.73) & 0.952 (0.83) & 0.958 (0.21) \\ \hline
4.0  &1,1 &  1.222  & 1.249 (2.21) & 1.250 (2.29) & 1.250 (2.29) \\
4.0  &1,2 &  0.905  & 0.911 (0.66) & 0.911 (0.66) & 0.896 (0.99)\\
4.0  &1,3 &  0.990  & 0.979 (1.11) & 0.979 (1.11) & 0.987 (0.30) \\
4.0  &1,4 &  0.989  & 0.971 (1.82) & 0.971 (1.82) & 0.985 (0.40)\\ \hline
\end{tabular}
\end{center}
\end{table}

\begin{table}
\begin{center}
Table 3. Ground state energies of pyracylene, fluoranthene and corannulene (in units of $t$)
from VMC and PQMC calculations.\\
\begin{tabular}{|c|cc|cc|cc|}\\ \hline 
\multicolumn{1}{|c|}{$U/t$}&\multicolumn{2}{c|}{Pyracylene}&\multicolumn{2}{c|}
{Fluoranthene} &\multicolumn{2}{c|}{Corannulene} \\ \hline 
 & $E_{VMC}$ & $E_{PQMC}$ &  $E_{VMC}$  &  $E_{PQMC}$ &  $E_{VMC}$  &  $E_{PQMC}$  \\ \hline 
0.5 & -17.713 & -17.725   & -20.558 & -20.565&-26.3058& -26317  \\
1.0 & $-16.116^{*}$ & -16.158  & -18.730 &-18.771&-24.007& -24.062 \\
2.0 & -13.235 & -13.406  & -15.422 &-15.581 &-19.8474 & -20.037\\ 
4.0 & -8.666 & -9.434  & -10.164 & -10.892&-13.158 & -14.063 \\ \hline
\end{tabular}
\end{center}
$^{*}$ exact energy obtained from full CI calculation in the singlet
subspace of dimension 2760615 is -16.13
\end{table}
\pagebreak
\clearpage

\begin{table}
\begin{center}
Table 4. Ground state energy of uniform and bond alternant $C_{60}$ for various values of $U$
from SPQMC calculations.\\
\begin{tabular}{|c|ccc|ccc|} \hline 
\multicolumn{1}{|c|}{$U/t$} & \multicolumn{3}{c|}{uniform}&
\multicolumn{3}{c|}{bond alternant} \\ \hline 
    & $E$ & $E/site$ & $E/bond$  & $E$  & $E/site$ & $E/bond$  \\ \hline 
1.0 & -79.0307  & -1.317 & -0.878  & -81.5973 & -1.360 & -0.907\\
2.0 & -66.6284  & -1.110 & -0.740  & -69.2986 & -1.155 & -0.770\\
3.0 & -56.0459  & -0.934 & -0.623  & -58.8326 & -0.981 & -0.654\\
4.0 & -47.3407  & -0.789 & -0.526  & -50.2902 & -0.836 & -0.559\\
5.0 & -40.4724  & -0.675 & -0.450  & -43.5544 & -0.726 & -0.484\\
6.0 & -35.6031  & -0.593 & -0.396  & -38.4948 & -0.642 & -0.428\\  \hline 
\end{tabular}
\end{center}
\end{table}

\begin{table}
\begin{center}
Table 5. On--site and nearest neighbour SPQMC charge--charge
correlations  ($<n_{i}n_{j}>$) and spin--spin correlations
($4<s_{i}^{z}s_{j}^{z}>$) and bond orders 
 of uniform and bond alternant $C_{60}$ for different values of $U/t$. \\
\begin{tabular}{|c|c|ccc|ccc|} \\ \hline 
\multicolumn{1}{|c|}{$U$} &\multicolumn{1}{c|}{}&
\multicolumn{3}{c|}{uniform}& \multicolumn{3}{c|}{bond alternant} \\ \hline 
 &  & h--h & h--p & diag  &  h--h& h--p& diag  \\ \hline 
1.0  & bo & 0.598 & 0.472  &---  & 0.708 & 0.408 & --- \\
  &  s-s  &-0.212  & -0.130  & 0.549 & -0.294 & -0.097 & 0.550 \\
  &  c-c  & 0.846  & 0.902  & 1.450 & 0.783 & 0.926 & 1.449 \\ \hline
2.0 &bo &  0.589 & 0.461  &---  & 0.700 & 0.396 & --- \\
  & s-s& -0.249 & -0.148  & 0.559 & -0.345 & -0.109 & 0.603 \\
  & c-c & 0.869 & 0.915  & 1.401 & 0.814 & 0.937 & 1.397 \\ \hline
3.0& bo  & 0.575 & 0.442  &---  & 0.687 & 0.374 &---  \\
  & s-s & -0.294 & -0.168  & 0.650 & -0.406 & -0.121 & 0.655 \\
 & c-c &0.888 & 0.928 & 1.350 & 0.840 & 0.947 & 1.345 \\ \hline
4.0 & bo& 0.554 & 0.415 &---  & 0.667 & 0.345 &---  \\
 & s-s &-0.350 & -0.192 & 0.702 & -0.473& -0.135 & 0.708 \\
 & c-c& 0.905 & 0.939 & 1.298 &0.863 &0.912 & 1.292 \\ \hline 
\end{tabular}
\end{center}
\end{table}
\pagebreak
\clearpage

\begin{table}
\begin{center}
Table 6. Difference in the bond orders of the h-h and h-p bonds
of $C_{60}$ from VMC and SPQMC calculations. \\
\begin{tabular}{|c|cc|cc|} \hline 
\multicolumn{1}{|c|}{$U/t$} & \multicolumn{2}{c|}{uniform}&
\multicolumn{2}{c|}{bond alternant} \\ \hline 
    & VMC & SPQMC & VMC  & SPQMC  \\ \hline 
1.0 & 0.126  & 0.126 & 0.482  & 0.300 \\
2.0 & 0.121  & 0.128 & 0.486  & 0.304 \\
3.0 & 0.121  & 0.133 & 0.436  & 0.313 \\
4.0 & 0.112  & 0.139 & 0.379  & 0.322 \\ \hline
\end{tabular}
\end{center}
\end{table}
\pagebreak

\pagebreak
\clearpage
\begin{center}
{\bf \underline{Figure Captions}} \\
\end{center}

\begin{description}
\item {\bf Figure 1.} Sample Ising  configurations (a) and (b)
 for the 4-site Hubbard 
ring with 4 time slices which are related by $C_{4}$ symmetry.
{\item {\bf Figure 2.}} Structures of (a) Pyracylene, (b) Fluoranthene and 
(c) Corannulene. 
{\item {\bf Figure 3.}}  (a) Structure of Radialene and 
(b) Pentagonal projection of $C_{60}$.
{\item {\bf Figure 4.}}  Energy/site of  (a) uniform and
(b) bond alternated $C_{60}$ for various values of $U/t$.
Open circles represent SPQMC data
while crosses are from the VMC calculation.
{\item {\bf Figure 5.}} Bond orders for the h-h and h-p bonds from SPQMC 
calculations (crosses and open triangles) and VMC calculations 
(open squares and open circles) for  (a) uniform and
(b) bond alternated $C_{60}$ for various values of $U/t$.
\pagebreak
\clearpage
{\item {\bf Figure 6.}} Inter-site spin-spin correlations of
(a) uniform and (b) bond alternant $C_{60}$ for
1-2 (filled squares), 1-11 (open circles), 1-3 (asterisks),
1-6 (crosses) and 1-7 (open triangles) site connectivities,
for different values of $U/t$.
{\item {\bf Figure 7.}} Inter-site charge-charge correlations of
(a) uniform and (b) bond alternant $C_{60}$ for
1-2 (open squares), 1-11 (open circles), 1-3 (asterisks),
1-6 (crosses) and 1-7 (open triangles) site connectivities,
for different values of $U/t$.
{\item {\bf Figure 8.}} $\rho_{\Gamma}$ for 
(a) uniform and (b) bond alternant $C_{60}$ for the
irreducible representations of the icosahedral point group,
 for $U/t = 1.0$ (asterisks), $U/t = 2.0$ (open circles),
$U/t = 3.0$ (triangles), $U/t = 4.0$ (crosses),
$U/t = 5.0$ (inverted triangles) and $U/t = 6.0$ (open squares).
{\item {\bf Figure 9.}} $S_{\Gamma}$ for 
(a) uniform and (b) bond alternant $C_{60}$ for the
irreducible representations of the icosahedral point group,
 for $U/t = 1.0$ (asterisks), $U/t = 2.0$ (open circles),
$U/t = 3.0$ (triangles), $U/t = 4.0$ (crosses),
$U/t = 5.0$ (inverted triangles) and $U/t = 6.0$ (open squares).

\end{description}

\end{document}